\documentclass{aastex62}

\usepackage{float}
\usepackage{upgreek}
\usepackage[shortlabels]{enumitem}
\usepackage[caption=false]{subfig}
\usepackage{graphbox}

\begin{document}
\title{Col-OSSOS: Compositional homogeneity of three Kuiper belt binaries} 

\author[0000-0001-8617-2425]{Micha$\ddot{\rm e}$l Marsset}
\affiliation{Department of Earth, Atmospheric and Planetary Sciences, MIT, 77 Massachusetts Avenue, Cambridge, MA 02139, USA}
\affiliation{Astrophysics Research Centre, Queen's University Belfast, Belfast BT7 1NN, United Kingdom}

\author[0000-0001-6680-6558]{Wesley C. Fraser}
\affiliation{NRC Herzberg Astronomy and Astrophysics, 5071 West Saanich Road, Victoria, BC, V9E 2E7, Canada}
\affiliation{Astrophysics Research Centre, Queen's University Belfast, Belfast BT7 1NN, United Kingdom}

\author[0000-0003-3257-4490]{ Michele T. Bannister}
\affiliation{School of Physical and Chemical Sciences --- Te Kura Mat\={u}, University of Canterbury, Private Bag 4800, Christchurch 8140, New Zealand}
\affiliation{Astrophysics Research Centre, Queen's University Belfast, Belfast BT7 1NN, United Kingdom}

\author[0000-0003-4365-1455]{Megan E. Schwamb}
\affiliation{Astrophysics Research Centre, Queen's University Belfast, Belfast BT7 1NN, United Kingdom}
\affiliation{Gemini Observatory, Northern Operations Center, 670 North A'ohoku Place, Hilo, HI 96720, USA}

\author[0000-0003-4797-5262]{Rosemary E. Pike}
\affiliation{Harvard \& Smithsonian Center for Astrophysics; 60 Garden Street, Cambridge, MA, 02138, USA}
\affiliation{Institute of Astronomy and Astrophysics, Academia Sinica; 11F of AS/NTU Astronomy-Mathematics Building, No.1, Sec. 4,, Roosevelt Rd Taipei 10617, Taiwan, R.O.C.}

\author[0000-0001-8821-5927]{Susan Benecchi}
\affiliation{Planetary Science Institute, 1700 East Fort Lowell, Suite 106, Tucson, AZ 85719, USA}

\author[0000-0001-7032-5255]{J. J. Kavelaars}
\affiliation{NRC Herzberg Astronomy and Astrophysics, 5071 West Saanich Road, Victoria, BC, V9E 2E7, Canada}
\affiliation{Department of Physics and Astronomy, University of Victoria, Elliott Building, 3800 Finnerty Rd, Victoria, BC V8P 5C2, Canada}


\author[0000-0003-4143-8589]{Mike Alexandersen} 
\affiliation{Harvard \& Smithsonian Center for Astrophysics; 60 Garden Street, Cambridge, MA, 02138, USA}
\affiliation{Institute of Astronomy and Astrophysics, Academia Sinica; 11F of AS/NTU Astronomy-Mathematics Building, No.1, Sec. 4,, Roosevelt Rd Taipei 10617, Taiwan, R.O.C.}

\author[0000-0001-7244-6069]{Ying-Tung Chen} 
\affiliation{Institute of Astronomy and Astrophysics, Academia Sinica; 11F of AS/NTU Astronomy-Mathematics Building, No.1, Sec. 4,, Roosevelt Rd Taipei 10617, Taiwan, R.O.C.}

\author[0000-0002-0283-2260]{Brett J. Gladman} 
\affiliation{Department of Physics and Astronomy, University of British Columbia, Vancouver, BC V6T 1Z1, Canada}

\author[0000-0001-8221-8406]{Stephen D. J. Gwyn} 
\affiliation{NRC Herzberg Astronomy and Astrophysics, 5071 West Saanich Road, Victoria, BC, V9E 2E7, Canada}

\author[0000-0003-0407-2266]{Jean-Marc Petit} 
\affiliation{Institut UTINAM UMR6213, CNRS, Univ. Bourgogne Franche-Comt\'e, OSU Theta F25000 Besan\c{c}on, France}

\author[0000-0001-8736-236X]{Kathryn Volk} 
\affiliation{Lunar and Planetary Laboratory, The University of Arizona, 1629 E. University Blvd., Tucson, AZ 85721, USA}



%
%
%
%
%
%

\email{michael.marsset@qub.ac.uk}

\begin{abstract}

The surface characterization of Trans-Neptunian Binaries (TNBs) is key to understanding 
the properties of the disk of planetesimals from which these objects formed. 
In the optical wavelengths, it has been demonstrated that most equal-sized component systems share similar colors, suggesting they have a similar composition. 
The color homogeneity of binary pairs contrasts with the overall diversity of colors in the Kuiper belt, which 
was interpreted as evidence that Trans-Neptunian Objects (TNOs) formed from a locally homogeneous and globally heterogeneous protoplanetary disk.  
In this paradigm, binary pairs must have formed early, before the dynamically hot TNOs were scattered out from their formation location. 
The latter inferences, however, relied on the assumption that the matching colors of the binary components imply matching composition. 
Here, we test this assumption by examining the component-resolved photometry of three TNBs found in the Outer Solar System Origins Survey: 
505447~(2013~SQ99), 511551~(2014~UD225) and 506121~(2016~BP81), 
across the visible and $J$-band near-infrared wavelength range. 
We report similar colors within 2$\upsigma$ for the binary pairs suggestive of similar reflectance spectra and hence surface composition.
This 
advocates for gravitational collapse of pebble clouds as a possible TNO formation route. 
We however stress that several similarly small TNOs, including at least one binary, have been shown to exhibit substantial spectral variability in the near-infrared, 
implying color equality of binary pairs is likely to be violated in some cases. 

\end{abstract}

\keywords {Kuiper belt: general  -- minor planets, asteroids: general  -- surveys}



\section{Introduction}
\label{sec:intro}


Since the discovery of Pluto's moon Charon \citep{Christy:1978dz}, more than a hundred binaries have been detected in the Kuiper belt\footnote{\url{http://www2.lowell.edu/users/grundy/tnbs/status.html}, 106 binary systems as of April 20 2020}, 
the population of planetesimals near and beyond the orbit of Neptune. 
Several mechanisms have been invoked for the origin of these systems, 
from mutual gravitational captures \citep{Goldreich:2002tb, Funato:2004vv, Schlichting:2008kc}, to disruptions of fast rotators \citep{Ortiz:2012cy}, 
and co-formation through gravitational collapses of pebble clouds in the primordial turbulent disk \citep{Johansen:2007cl, Nesvorny:2010da}. 
Each formation mechanism is expected to result in different physical properties, such as size, size ratio, orbit pole orientation, and (dis)similarity in composition, of the Trans-Neptunian Binaries (TNBs) components. 
Simulations of co-formation by gravitational collapses currently provide the best match to the observed properties of the equal-size TNBs \citep{Nesvorny:2019, Robinson:2020}, including their prevalence (\citealt{Noll:2008bj, Fraser:2017kh}; although maybe not at  small sizes $H_r>7.5$, \citealt{pike:2020}), their broad inclination distribution, and their predominantly prograde orbits \citep{Grundy2019}. 
This scenario also explains the high fraction of widely-separated binaries in the dynamically quiescent `cold' (low inclination and eccentricity) population of the Kuiper belt, compared to the dynamically `hot' component where wide binaries would have been disrupted by Neptune scattering \citep{Parker:2010, Nesvorny:2019b}. 
In this scenario, the TNBs would share the same composition, having accreted in the same environment. 

In the optical wavelengths, \citet{Benecchi:2009jo} found that the components of similar-sized TNBs share the same colors, 
and interpreted this as evidence that they have similar composition. 
The similar colors of the pair members further contrast with the variety of colors among the overall population of Trans-Neptunian Objects (TNOs; e.g., \citealt{Tegler:1998jl, Tegler:2000fx, Tegler:2003jx, Tegler:2003is, Tegler:2008vj, Tegler:2016ep, Peixinho:2003jn, Peixinho:2012bt, Peixinho:2015bw, Fraser:2012cs, Lacerda:2014hw, Fraser:2015cx, Pike:2017gf, Wong:2017ck, Schwamb:2019, Marsset:2019, Thirouin:2019, Alvarez_Candal_2019}), suggesting these objects formed from a locally homogeneous and globally heterogeneous protoplanetary disk. 
This inference however relies on the assumption that matching optical colors imply matching composition. 
However, optical colors are not sufficient to infer compositional (dis)similarities of the binary components, because different ices and chemical elements in the outer Solar System share similar optical colors (e.g., \citealt{Barucci:2011}), 
and many of these elements only distinctively reveal themselves outside of optical wavelengths. 
Measuring composition would normally require measuring albedo and reflectance spectra of the components across the full visible and near-infrared spectral ranges. 
As most TNOs are too faint for spectroscopy, one must instead rely on what near simultaneous multi-band photometry reveal as a proxy.

In this short paper, we present the colors of three TNBs across the visible and $J$-band wavelength range and test the assumption of an identical bulk composition for the primary and secondary components of each binary pair.
To do so, we reanalyze recent near-simultaneous $g$, $r$, $z$ and $J$-band photometric data collected with 
Gemini North through the Colours of the Outer Solar System Origins Survey (Col-OSSOS; \citealt{Schwamb:2019}), in order to search for new TNBs and explore the spectral (dis)similarities of their components across these spectral bands. 

\section{Search and characterization of candidate binaries}
\label{sec:obs}

\subsection{Observations}


Photometric measurements presented in this paper were collected from the ground through Col-OSSOS on the 8.1~m Frederick C. Gillett Gemini-North Telescope on Maunakea. 
Col-OSSOS acquires near-simultaneous \emph{g}, \emph{r} and \emph{J} -- and, for a subsample of objects, z \citep{Pike:2017gf} -- photometry of a magnitude-limited subset of the Outer Solar System Origins Survey (OSSOS; \citealt{Bannister:2016cp, Bannister:2018ha}) sample with \emph{r}$<$23.6 \citep{Schwamb:2019}. 
Measurements were acquired in a \emph{rg(z)J(z)gr} sequence to monitor any lightcurve effect, using the Gemini Multi-Object Spectrograph (GMOS, \citealt{Hook:2004fq}) for measurements in the optical wavelengths and the Near InfraRed Imager and Spectrometer (NIRI, \citealt{Hodapp:2003ko}) for those in the $J$-band. 
Targets were tracked sidereally, meaning that stars appear point-like on the images while the TNO is elongated in the direction of its projected motion on the sky. 
Exposures were limited to 300~s in the optical and 120~s in $J$-band in order to minimize trailing losses and to mitigate the sky background. 
The telescope was dithered between consecutive exposures in the same filter to account for pixel-to-pixel sensitivity variations. 
Observing circumstances for our targets can be found in \citet{Pike:2017gf} and  \citet{Schwamb:2019}. 


\subsection{Data reduction}


The data reduction and photometric extraction of the Gemini data were performed as described in \citet{Schwamb:2019}. 
Here we summarize the main steps.
Processing of the GMOS data was achieved with the Gemini-IRAF package. 
We first used the bias images acquired as part of the GMOS calibration plan to remove the bias offset from the science frames. 
Master sky flats were produced from an average of the science frames with sources masked, and used to flatten these frames. 
Finally, the GMOS CCD chips were mosaicked into a single extension. 

For the NIRI images, flat-fielding was performed with a master flat built from the Gemini facility calibration unit flats. 
We produced one sky frame for each individual science image from a rolling average of the 15 temporally closest science images with sources masked, in order to account for temporal sky variation. 
Images acquired at similar dither position to the image to sky-subtract were not used to compute the average sky frame. 
Bad pixel maps were created from the individual dark exposures, using the longest exposures to flag dead and low-sensitive pixels, and the shortest ones to identify hot pixels. 
These maps were then combined with the NIRI static bad pixel map provided in the Gemini IRAF package. 
Individual sky-subtracted images were aligned sidereally using multiple star centroids in order to create a deep stacked image of point-like stars that can be used to compute the mean point-spread function (PSF) of the image. 
A second stack was produced shifting the frames at the TNO's on-sky velocity derived from the MPC Minor Planet and Comet Ephemeris Service\footnote{\url{http://www.minorplanetcenter.net/iau/MPEph/MPEph.html}. Orbits, and thus on-sky velocities, are those published in \citet{Bannister:2018ha}}. 
Finally, Cosmic ray rejection was performed on the stacked images using the Python implementation of the L.A.Cosmic (Laplacian Cosmic Ray Identification) algorithm\footnote{\url{http://www.astro.yale.edu/dokkum/lacosmic/}}.

\subsection{Trailed Point-Spread Function fitting}




A binary search was performed using the complete set of GMOS and NIRI images acquired for each target in the Col-OSSOS survey. 
We calculated and subtracted the Trailed Point-Spread Function (TSF) of each image with the TRIPPy software package \citep{Fraser:2016bd}. 
The TSF is the equivalent of the PSF for a moving object. 
The details of the method used to model the TSF are provided in \citet{Fraser:2016bd}. 
We describe it briefly here. 
First, the PSF of each GMOS image and NIRI stack of images was measured by fitting a Moffat profile to each stellar (point-like) source on the frame. 
The best-fit solution was derived by finding the minimum $\chi^2$ residual between the mean stellar profile and the modeled Moffat profile. 
Next, the mean Moffat profile was subtracted from each stellar source to create a mean residual image, which was then added to the mean Moffat profile.  
The resulting Moffat+residual profile was convolved with a line having length and angle that best describes the known on-image rate of motion of the TNO to produce the TSF. 
The TSF was then fitted to the TNO image using the emcee Python implementation of \citet{ForemanMackey:2013, Foreman-Mackey:2019}'s Markov Chain Monte Carlo (MCMC) ensemble sampler. 
A total of 20 walkers with three free parameters: the (x,y) positions and flux of the TNO, were used to find the minimum $\chi^2$ residual between the model and the image. 
Once a solution was found, uncertainties on the parameters were calculated as the 1-$\upsigma$ deviation of the walkers across an extra 100 iteration steps.  
Finally, we built stacks of the residual images acquired in the same filter for each object in our dataset in order to increase the Signal-to-Noise Ratio (SNR) of the residual, and to search for very faint companions. 


\subsection{Analysis of the residuals}

\begin{figure}[h!]
\centering
\includegraphics[angle=0, width=0.3\linewidth, trim=0cm 1.57cm 0cm 0cm, clip]{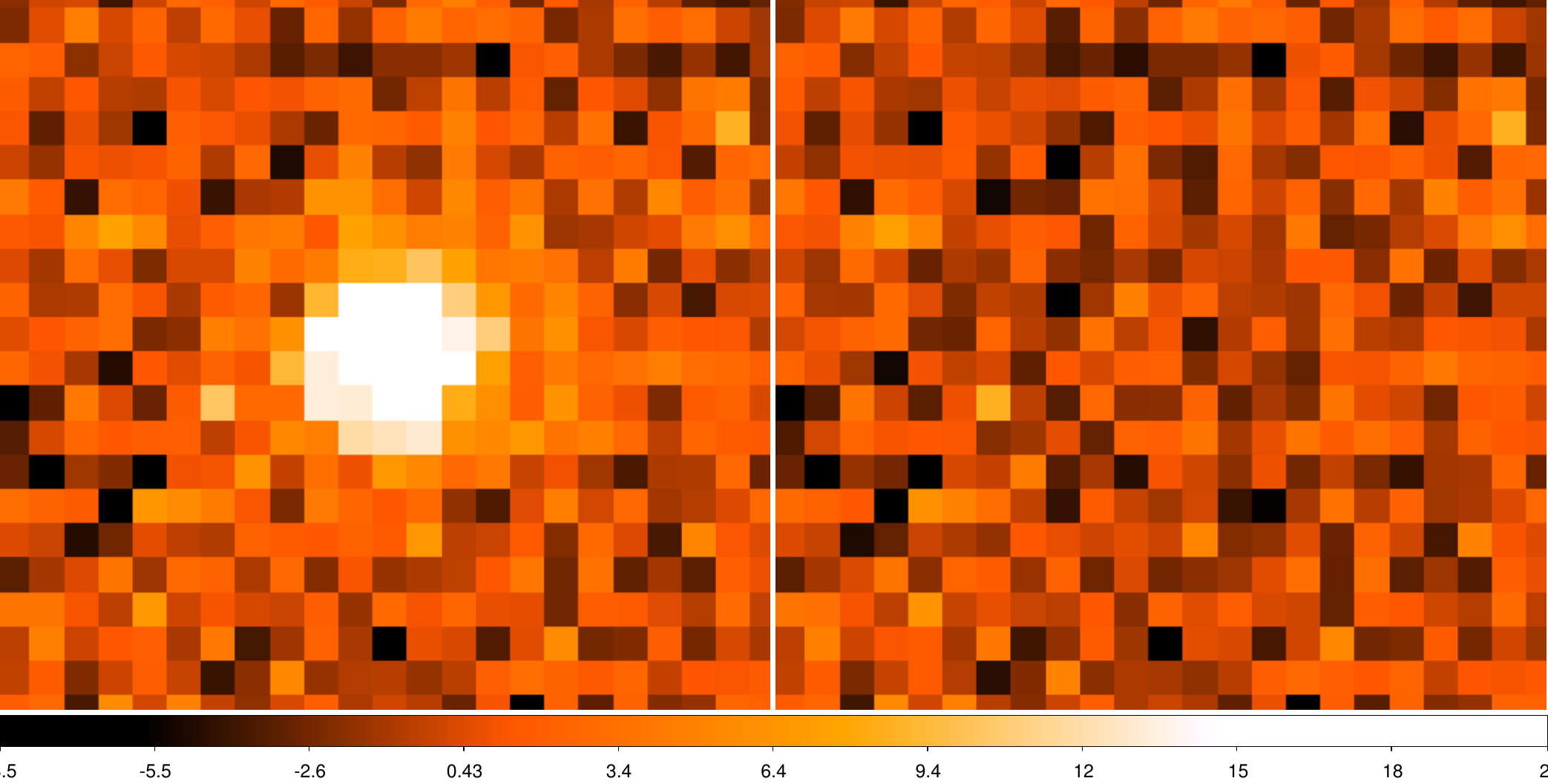}
 \caption{ Example of single-TSF subtraction of a non-binary TNO with the TRIPPy software \citep{Fraser:2016bd}. {\it Left:} Original NIRI stacked image of 2013~GQ137 (OSSOS ID o3e21). {\it Right:} Single-TSF subtracted image. The residuals are indistinguishable from the background noise. Image scales, contrast, and color bars are the same for the two images.}
\label{fig:colossos_single}
\end{figure}

Three distinct kinds of residuals are revealed by our analysis. 
The first kind is consistent with the TNO being a single source. 
In that case, the residuals are indistinguishable from the background noise, i.e., the pixel values in the region of the subtracted TNO follow a Poisson distribution (Fig.~\ref{fig:colossos_single}). 
In the case of a double source, two categories of residuals are observed. 
When the binary has roughly similar-sized components, a `butterfly' pattern is revealed as a result of subtracting a single TSF to a double-peak signal. 
When the brightness contrast between the two components is large, the primary is efficiently fitted and removed by the TSF, and a round, secondary source is revealed in the residuals. 
Double-TSF fitting on these double sources result in residuals that are indistinguishable from the background. 

\begin{figure}[h!]
\centering
\includegraphics[angle=0, width=0.45\linewidth, trim=5cm 1cm 5cm 0cm, clip]{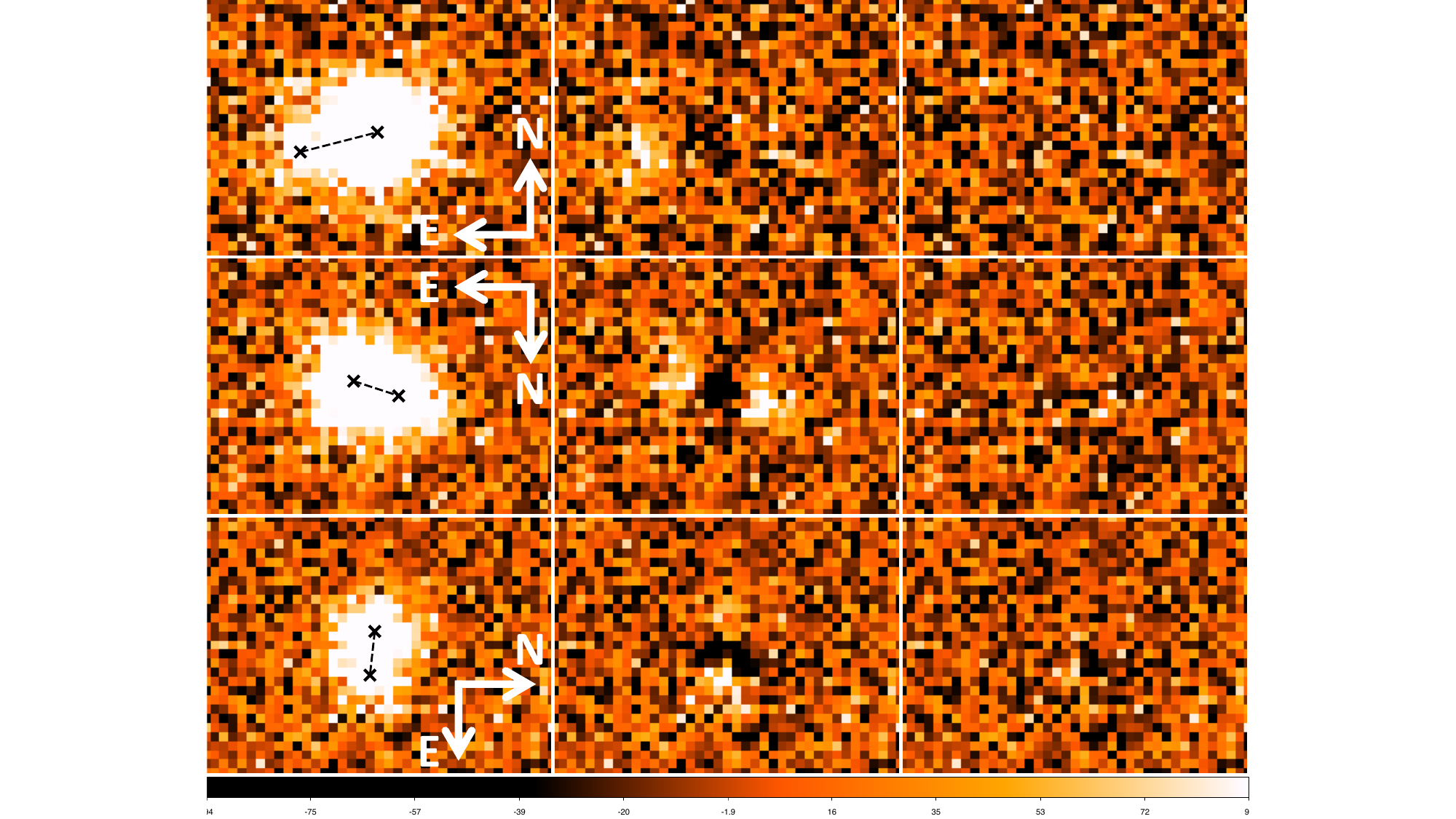}
 \caption{ {\it Left column:} GMOS images acquired on Gemini North for 511551 (2014~UD225) ({\it top}), 506121 (2016~BP81) ({\it middle}) and 505447~(2013~SQ99) ({\it bottom}). 
 Black crosses indicate the location of the two components's centroids as determined by the TSF fitting procedure. The white arrows indicate the equatorial North (N) and East (E) directions. {\it Middle column:} Single-TSF subtraction reveals the binarity of the three TNOs. {\it Right column:} Double-TSF subtraction efficiently removes the two binary components, with a residual consistent with the background. Image scales, contrast, and color bars are the same for all images.}
\label{fig:colossos_binaries}
\end{figure}

Our analysis allowed the identification of three TNBs among 75 objects searched in the Col-OSSOS dataset: 511551~(2014~UD225), 506121~(2016~BP81) and 505447~(2013~SQ99). 
Images of the binaries are shown in Fig.~\ref{fig:colossos_binaries}\footnote{From 
top to bottom, those correspond to the r-band image mrgN20140826S0221.fits of 511551~(2014~UD225), 
the r-band image mrgN20140826S0174.fits of 506121~(2016~BP81),
and the g-band image mrgN20140822S0267.fits of 505447~(2013~SQ99) (filenames listed in the Gemini Observatory Archive).}
and their parameters and observing circumstances are provided in Table~\ref{tab:objects}. 
All three objects come from the first (2014B) semester of Col-OSSOS observations and their discovery has been previously reported by \citet{Fraser:2017kh}. 
No other TNB could be identified in the subsequent Col-OSSOS semesters. 
We attribute this to the worse seeing conditions experienced during later semesters with respect to our first semester of observations, where seeing was between $0".4-0".5$. 

In the case of 511551, the components's brightness ratio is large and single TSF-fitting efficiently removed the primary and revealed the presence of a smaller companion. 
In the case of 506121 and 505447, the components have roughly similar brightness and TSF-subtraction produced a `butterfly-shaped' residual (Fig.~\ref{fig:colossos_binaries}). 
The physical properties -- angular separation and components's brightness ratio -- of the three binaries are investigated in the following Section.

\startlongtable
\begin{deluxetable}{lllcccclllllcll}
\tabletypesize{ \scriptsize}
\tablecaption{\label{tab:objects} Orbital Parameters and Observing Circumstances for the Three Binaries. }
\tablehead{\colhead{Target} & \colhead{MPC} &  \colhead{OSSOS} &  \colhead{a (au)} &  \colhead{e} &  \colhead{inc ($\degr$)} & \multicolumn{1}{c}{ Orbital } & \colhead{Mean $m_{\rm r}$} & \colhead{H${\rm _r}$} & \colhead{$\Delta$ (au)} & \colhead{$r_{\rm H}$ (au)} &  \colhead{$\alpha$ ($\degr$)} &  \colhead{ Average } &  \colhead{Mean} \\ 
\colhead{Number} & \colhead{Designation} & \colhead{ID} & \colhead{ } & \colhead{  } & \colhead{ } & \multicolumn{1}{c}{ Classification } & \colhead{ (SDSS) } & \colhead{  } & \colhead{  } & \colhead{  } & \colhead{  } & \colhead{ Air Mass } &  \colhead{MJD} }
\startdata
511551 & 2014~UD225 & o4h45 & 43.36 & 0.130 & 3.66 & cold classical & 22.98$\pm$0.05 & 6.55 & 43.68 & 44.29 & 1.05 & 1.03 & 56895.554093 \\  
--"--  & --"-- & --"-- & --"-- & --"-- & --"-- & --"-- & 23.17$\pm$0.04 & --"-- & 43.65 & --"-- & 1.02 & 1.38 & 56897.438239 \\  
506121 & 2016~BP81 & o3l39 & 43.68 & 0.076 & 4.18 & 7:4 resonator & 22.83$\pm$0.11 & 6.58 & 41.81 & 42.54 & 0.96 & 1.13 & 56895.471738 \\ 
505447 & 2013~SQ99 & o3l76 & 44.15 & 0.093 & 3.47 & cold classical & 23.17$\pm$0.04 & 6.45 & 46.60 & 47.34 & 0.85 & 1.05 & 56891.544967 \\ 
  \enddata
\tablenotetext{}{Geometric parameters and derived H${\rm _r}$ are reported for the time of the Col-OSSOS observation.} 
\tablenotetext{}{$r_{\rm H}$: heliocentric range, $\Delta$: observer range, $\alpha$: phase angle.}
\tablenotetext{}{Orbital classifications from \citet{Bannister:2018ha}.}
\tablenotetext{}{511551~(2014~UD225) was observed at two distinct epochs in Col-OSSOS. Only one image from the second epoch shows evidence of binarity and was used in our analysis.}
\end{deluxetable}

\subsection{Component-resolved photometry}
\label{sec:measures}

The angular separation and components's brightness ratio of the binaries were derived through double TSF fitting using TRIPPy. 
We began by estimating the center position and brightness ratio of the two binary components by eye using both the original images and the single-TSF subtracted ones. 
The best fit solution was then derived for each observation through automated MCMC fitting of the sky-subtracted image. 
We used a total of 30 walkers, each performing a total number of a few hundred steps in the 6-parameter space defined by the position and flux of the primary (x$_1$, y$_1$ and f$_1$) and secondary (x$_2$, y$_2$ and f$_2$) components of the binary. 
At each step, a standard $\chi^2$ formula was used to evaluate the goodness of the fit between the synthetic double TSF image and the science image,  
using an uncertainty map to mitigate the influence of bad pixels when evaluating the $\chi^2$.   
Once the walkers converged towards a stable solution, an additional 100 steps per walker were run and recorded. 
We adopted as best-fit x$_1$, y$_1$, f$_1$, x$_2$, y$_2$ and f$_2$ parameters the median value of these parameters across the 30$\times$100 steps, and the uncertainty as their 1-$\upsigma$ deviation with respect to the median value.  
Results of our MCMC fitting analysis for individual images are reported in Table~\ref{tab:binary_properties1}. 
Adopted physical properties of the binaries -- including their component separation and brightness ratio -- are provided in Table~\ref{tab:binary_properties2}. 

\begin{figure}[ht]
\centering
\includegraphics[trim=0mm 0mm 0mm 0mm, clip, width=0.7\textwidth]{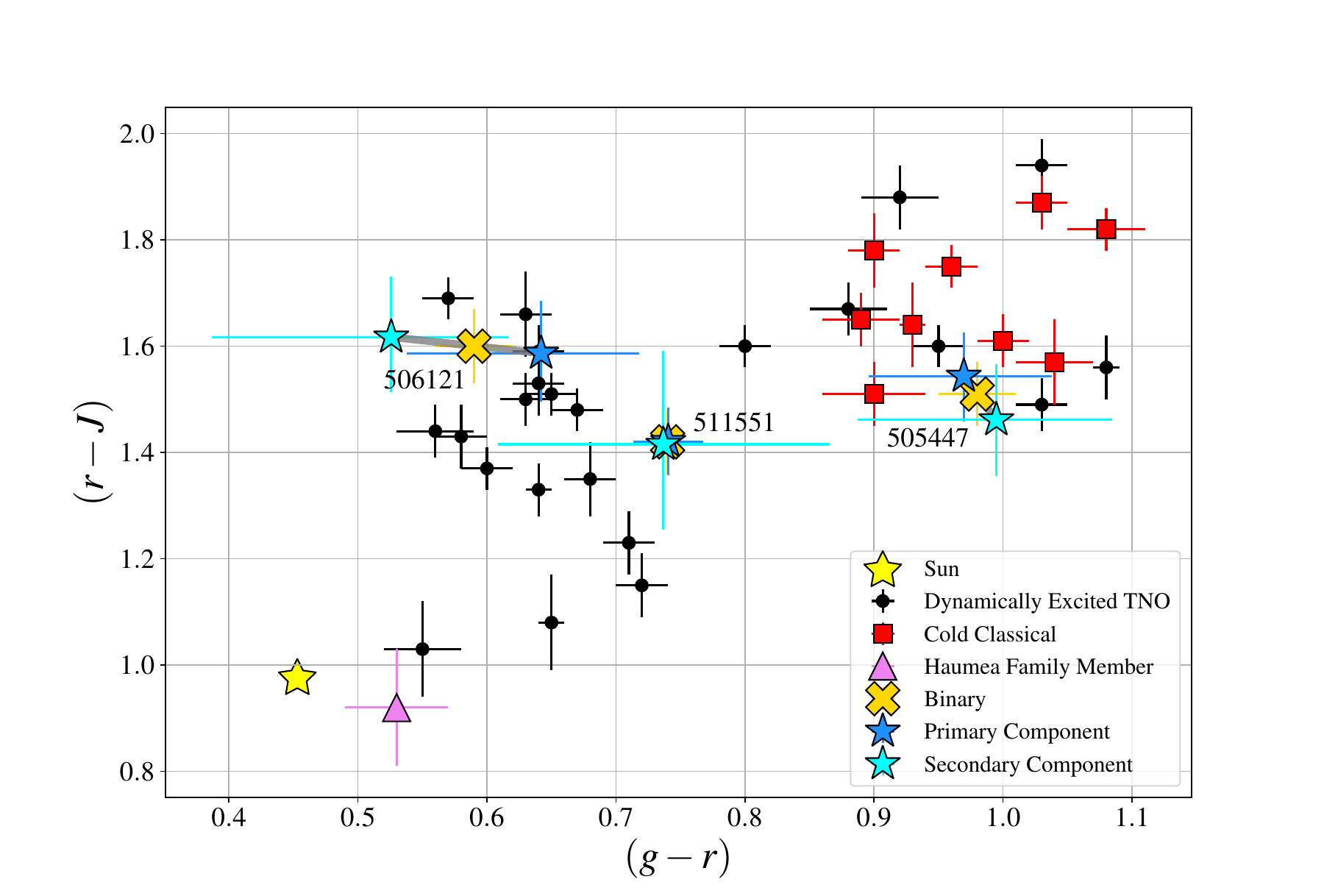}
\caption{ $(r-J)$ versus $(g-r)$ color-color plot of the Col-OSSOS dataset from \citet{Schwamb:2019}. 
The three binaries are marked by yellow crosses and their 
individual components by blue (primary component) and cyan (secondary component) stars. 
The IAU number of each binary is indicated next to the corresponding data point. 
Measurements of two pair members belonging to the same system are linked by a grey line. 
Colors of the 511551~(2014~UD225) system and its components are almost equal, making them hard to distinguish on the figure.  
Other objects from the Col-OSSOS dataset include singletons located on dynamically-excited orbits (black dots) and classicals objects with i$<$5$\degr$ (red squares). 
 The object 2013~UQ15 (magenta triangle) is bluer than the Sun in $(r-J)$ and 
dynamically consistent with the Haumea family.
 The solar color, with $(g-r)=0.45$ and $(r-J)=0.98$, is shown by the yellow star.} 
\label{fig:colorcolor}
\end{figure}

Next, we computed the component-resolved colors of the binaries by 
combining their measured $f_2$/ $f_1$ component brightness ratio with the magnitudes of the binaries converted to the SDSS photometric system reported in \citet{Schwamb:2019}. 
In each image, $f_2$/ $f_1$ was calculated directly from the amplitude coefficients used to scale the TRIPPy TSFs to match image brightnesses. 
Fluxes were estimated from planted sources in blank images, and from both, mean colours were measured in $(g-r)$, $(r-z)$ and $(r-J)$. 
Quoted uncertainties include the brightness ratio ranges from the fits, the Poisson statistics uncertainties, the zeropoint uncertainties, the background uncertainties and the aperture correction uncertainties. 
The derived $(g-r)$, $(r-z)$ and $(r-J)$ colors of the binaries and their associated uncertainties are reported in Table~\ref{tab:binary_photometry} and compared to the population of singletons in Fig.~\ref{fig:colorcolor}.

\begin{deluxetable}{cccccccccc}[h!]
\tabletypesize{ \scriptsize}
\tablecaption{ \label{tab:binary_properties1}
Results of double-TSF fitting. }
\tablehead{\multicolumn{1}{c}{ Object } & \multicolumn{1}{c}{ Image } & \multicolumn{1}{c}{ Filter } & \multicolumn{1}{c}{ MJD } & \multicolumn{1}{c}{ FWHM } &  \multicolumn{3}{c}{ Separation (") } & \colhead{ Brightness Ratio$^{**}$ } \\ 
\multicolumn{1}{l}{  } & \multicolumn{1}{c}{ } & \multicolumn{1}{l}{ } & \multicolumn{1}{l}{ } & \colhead{ (") } & \colhead{ $\delta$\,R.A.\,cos(Dec.)$^*$ } & \colhead{ $\delta$\,Dec.$^*$ } & \colhead{ Distance } & \colhead{ $\left(\frac{f_2}{f_1}\right)$ } } 
\startdata
505447                       & N20140822S0260.fits  & $z_{\rm \_G0304}$ & 56891.5052495 & 0.42 & ${-0.41}\pm{0.02}$              & ${0.06}\pm{0.02}$         & ${0.41}\pm{0.02}$        & ${0.65}^{+0.13}_{-0.09}$  \\
--"--                             & N20140822S0261.fits  & $z_{\rm \_G0304}$ & 56891.5096761 & 0.33 & ${-0.38}^{+0.02}_{-0.01}$    & ${0.04}\pm{0.02}$         & ${0.38}\pm{0.01}$        & ${0.54}^{+0.06}_{-0.07}$ \\
--"--                             & N20140822S0262.fits  & $z_{\rm \_G0304}$ & 56891.5140979 & 0.36 & ${-0.37}\pm{0.01}$              & ${0.02}^{+0.02}_{-0.01}$   & ${0.37}^{+0.01}_{-0.02}$ & ${0.61}^{+0.06}_{-0.05}$ \\
--"--                             & N20140822S0263.fits  & $z_{\rm \_G0304}$ & 56891.5185195 & 0.36 & ${-0.37}\pm{0.02}$              & ${0.02}^{+0.01}_{-0.02}$    & ${0.37}\pm{0.02}$        & ${0.60}^{+0.05}_{-0.04}$ \\
--"--                             & N20140822S0264.fits  & $r_{\rm \_G0303}$ & 56891.5229899 & 0.39 & ${-0.37}\pm{0.01}$              & ${0.02}\pm{0.01}$         & ${0.37}\pm{0.01}$        & ${0.71}\pm{0.06}$  \\
--"--                             & N20140822S0265.fits  & $g_{\rm \_G0301}$ & 56891.5274939 & 0.41 & ${-0.38}\pm{0.01}$             & ${0.03}^{+0.01}_{-0.02}$  & ${0.38}\pm{0.01}$        & ${0.80}^{+0.06}_{-0.08}$ \\
--"--                             & N20140822S0266.fits  & $g_{\rm \_G0301}$ & 56891.5319220 & 0.41 & ${-0.38}\pm{0.01}$             & ${0.03}\pm{0.02}$           & ${0.38}\pm{0.01}$        & ${0.59}^{+0.06}_{-0.05}$ \\
--"--                             & N20140822S0267.fits  & $g_{\rm \_G0301}$ & 56891.5363661 & 0.45 & ${-0.34}\pm{0.02}$             & ${0.04}\pm{0.02}$            & ${0.35}\pm{0.02}$        & ${0.70}^{+0.10}_{-0.12}$ \\
--"--                             & O13BL3TF.fits              & J                              & 56891.5864204 & 0.44 & ${-0.37}^{+0.03}_{-0.02}$  & ${0.02}\pm{0.02}$         & ${0.38}^{+0.03}_{-0.02}$ & ${0.66}^{+0.08}_{-0.06}$ \\
511551                       & N20140826S0184.fits  & $r_{\rm \_G0303}$ & 56895.5319392 & 0.49 & ${0.66}^{+0.04}_{-0.05}$ & ${0.12}^{+0.06}_{-0.04}$ & ${0.69}\pm{0.04}$ & ${0.11}\pm{0.02}$ \\
--"--                             & N20140826S0185.fits  & $g_{\rm \_G0301}$ & 56895.5364416 & 0.54 & ${0.72}\pm{0.05}$ & ${0.30}\pm{0.04}$ & ${0.80}^{+0.04}_{-0.05}$ & ${0.12}\pm{0.02}$ \\
--"--                             & N20140826S0217.fits  & $g_{\rm \_G0301}$ & 56895.6004712 & 0.53 & ${0.65}^{+0.03}_{-0.04}$  & ${0.15}^{+0.03}_{-0.04}$  & ${0.69}^{+0.03}_{-0.04}$ & ${0.14}^{+0.02}_{-0.01}$ \\
--"--                             & N20140826S0218.fits  & $g_{\rm \_G0301}$ & 56895.6048958 & 0.52 & ${0.67}^{+0.03}_{-0.04}$  & ${0.16}\pm{0.03}$         & ${0.71}^{+0.03}_{-0.04}$ & ${0.14}\pm{0.01}$ \\
--"--                             & N20140826S0219.fits  & $g_{\rm \_G0301}$ & 56895.6093289 & 0.53 & ${0.71}\pm{0.03}$         & ${0.16}\pm{0.02}$         & ${0.75}\pm{0.03}$        & ${0.15}\pm{0.01}$ \\
--"--                             & N20140826S0220.fits  & $r_{\rm \_G0303}$ & 56895.6138240 & 0.53 & ${0.64}^{+0.03}_{-0.02}$  & ${0.14}\pm{0.02}$         & ${0.67}^{+0.03}_{-0.02}$ & ${0.14}\pm{0.01}$ \\
--"--                             & N20140826S0221.fits  & $r_{\rm \_G0303}$ & 56895.6182511 & 0.53 & ${0.58}\pm{0.02}$ & ${0.16}^{+0.01}_{-0.02}$ & ${0.62}\pm{0.02}$ & ${0.18}\pm{0.01}$ \\
--"--                             & Col3N03.fits                 & J                            & 56895.5762472 & 0.55  & ${0.61}\pm{0.03}$           & ${0.10}\pm{0.04}$         & ${0.64}\pm{0.03}$        & ${0.14}\pm{0.02}$ \\
--"--                             & N20140828S0128.fits  & $r_{\rm \_G0303}$ & 56897.4382394 & 0.50 & ${0.64}\pm{0.04}$         & ${0.12}^{+0.02}_{-0.03}$ & ${0.65}\pm{0.04}$        & ${0.14}\pm{0.02}$ \\
506121                        & O13BL3RB.fits              & J                             & 56895.4564207 & 0.62 & ${-0.43}\pm{0.03}$         & ${-0.11}\pm{0.02}$         & ${0.44}\pm{0.04}$ & ${0.79}^{+0.06}_{-0.08}$ \\
--"--                              & N20140826S0173.fits  & $g_{\rm \_G0301}$ & 56895.4837177 & 0.59 & ${-0.34}\pm{0.02}$         & ${-0.15}^{+0.01}_{-0.02}$  & ${0.37}\pm{0.02}$        & ${0.85}^{+0.09}_{-0.15}$ \\
--"--                              & N20140826S0174.fits  & $r_{\rm \_G0303}$ & 56895.4870562 & 0.50 & ${-0.36}\pm{0.01}$          & ${-0.11}\pm{0.01}$ & ${0.38}\pm{0.01}$ & ${0.77}^{+0.07}_{-0.06}$ \\
\enddata
\tablenotetext{}{ ${}^{*}$Measured position of the brightest component minus position of the faintest one. }
\tablenotetext{}{ ${}^{**}$Measured flux of the faintest component divided by the measured flux of the brightest one. }
\end{deluxetable}

\begin{deluxetable}{ccccccc}[h!]
\tabletypesize{ \scriptsize}
\tablecaption{ \label{tab:binary_properties2}
Mean Physical Properties of the Binaries. }
\tablehead{\multicolumn{1}{c}{ Object } & \multicolumn{1}{c}{ Filter } & \colhead{ Number of } & \colhead{Separation } & \colhead{Separation } & \colhead{ Brightness Ratio } \\ 
\multicolumn{1}{l}{  } & \multicolumn{1}{l}{ } & \colhead{ images } & \colhead{ (") } & \colhead{ ($10^3$km) } & \colhead{ $\left(\frac{f_2}{f_1}\right)$ } } 
\startdata
505447         & g$_{\rm \_G0301}$     & 3 & ${0.37}^{+0.02}_{-0.01}$ & $12.5^{+0.7}_{-0.3}$   & ${0.70}^{+0.08}_{-0.09}$ \\
--"-- 		    & r$_{\rm \_G0303}$     & 1 & ${0.37}\pm{0.01}$            & $12.7\pm0.3$              & ${0.71}\pm{0.06}$ \\
--"-- 		   & z$_{\rm \_G0304}$    & 4 & ${0.38}\pm{0.02}$            & $12.9\pm0.7$              & ${0.60}^{+0.08}_{-0.06}$ \\
--"-- 		   & J                                 & 1 & ${0.38}^{+0.03}_{-0.02}$ & $12.9^{+1.0}_{-0.7}$    & ${0.66}^{+0.08}_{-0.06}$ \\              
511551        & g$_{\rm \_G0301}$    & 4 & ${0.74}^{+0.03}_{-0.04}$  & $23.5^{+1.0}_{-1.3}$    & ${0.14}^{+0.02}_{-0.01}$ \\
--"-- 		   & r$_{\rm \_G0303}$     & 4 & ${0.66}\pm{0.03}$            & $21.0\pm1.0$               & ${0.14}\pm{0.01}$ \\
--"-- 		   & J                                 & 1 & ${0.64}\pm{0.03}$            & $20.4\pm0.6$                & ${0.14}\pm{0.02}$ \\
506121        & g$_{\rm \_G0301}$     & 1 & ${0.37}\pm{0.02}$            & $11.2\pm0.6$                & ${0.85}^{+0.09}_{-0.15}$ \\
--"-- 		   & r$_{\rm \_G0303}$     & 1 & $0.37\pm0.01$                 &  $11.2\pm0.3$               & ${0.77}^{+0.07}_{-0.06}$ \\
--"-- 		   & J                                 & 1 & ${0.44}\pm{0.04}$              & $13.1\pm1.2$                & ${0.79}^{+0.06}_{-0.08}$ \\
\enddata
\end{deluxetable}

Next, all color measurements were converted to spectral slope (s), defined as the percentage increase in reflectance per ${\rm 10^3 \AA}$ change in wavelength normalised to 550~nm, 
using the Synphot tool in the STSDAS software package\footnote{\url{www.stsci.edu/institute/software\_hardware/stsdas}} \citep{Lim:2015}. 
Specifically, this was achieved by forward modelling a Solar spectrum convolved with a set of linear spectra, 
with spectral slope varying from neutral (${\rm 0\%/(10^3 \AA})$) to very red (${\rm 50\%/(10^3 \AA})$) with steps of ${\rm 0.1\%/(10^3 \AA})$. 
We used the solar reference spectrum provided by the Hubble Space Telescope (HST) as part of the Synphot supplementary files. 
The array of models was then passed through Synphot to estimate the corresponding colors in the various combinations of filters relevant for this work.

\begin{figure}[ht]
\centering
\begin{minipage}[b]{0.4\linewidth}
\centering
\includegraphics[trim=0mm 0mm 0mm 0mm, clip, width=\textwidth]{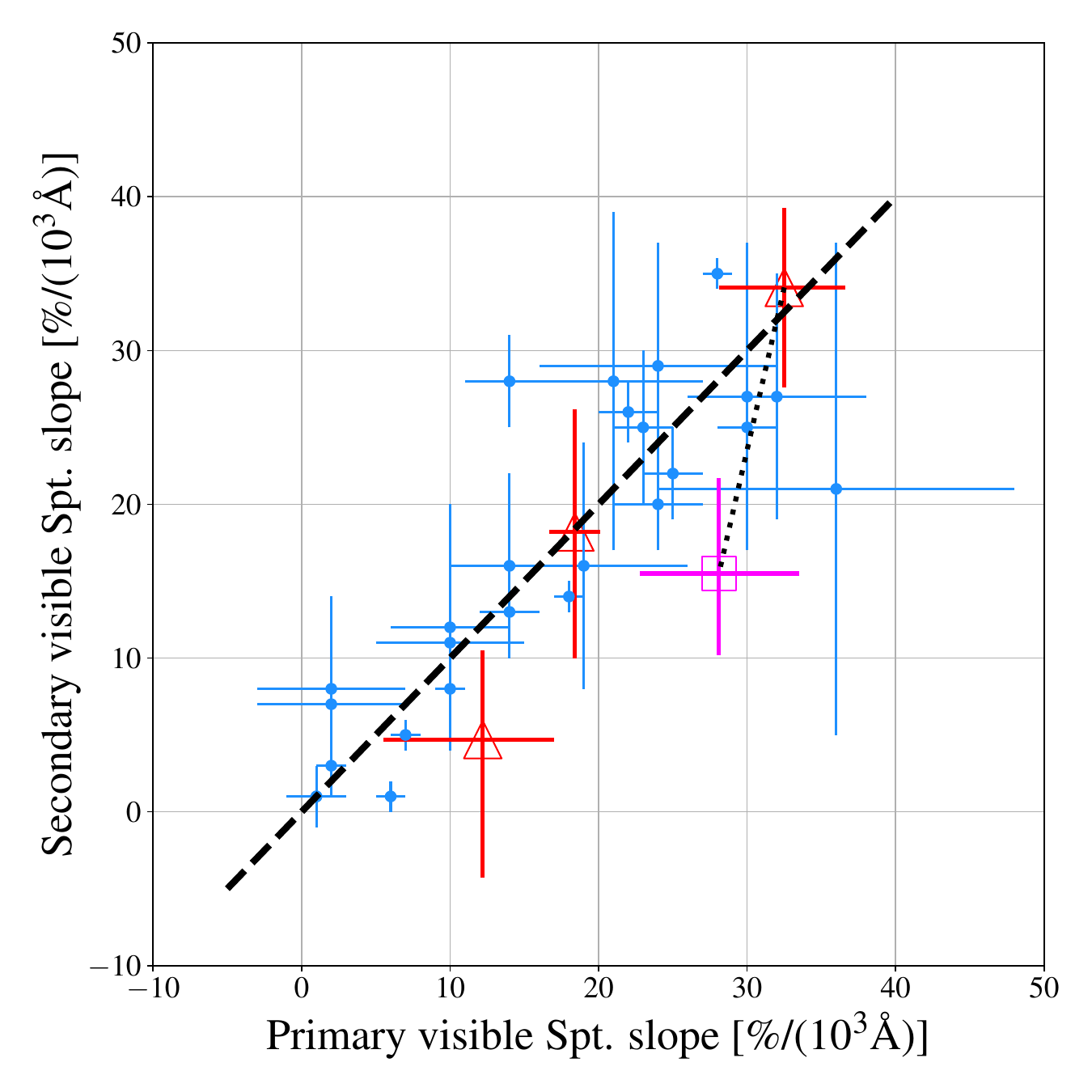}
\end{minipage}
\begin{minipage}[b]{0.4\linewidth}
\centering
\includegraphics[trim=0mm 0mm 0mm 0mm, clip, width=\textwidth]{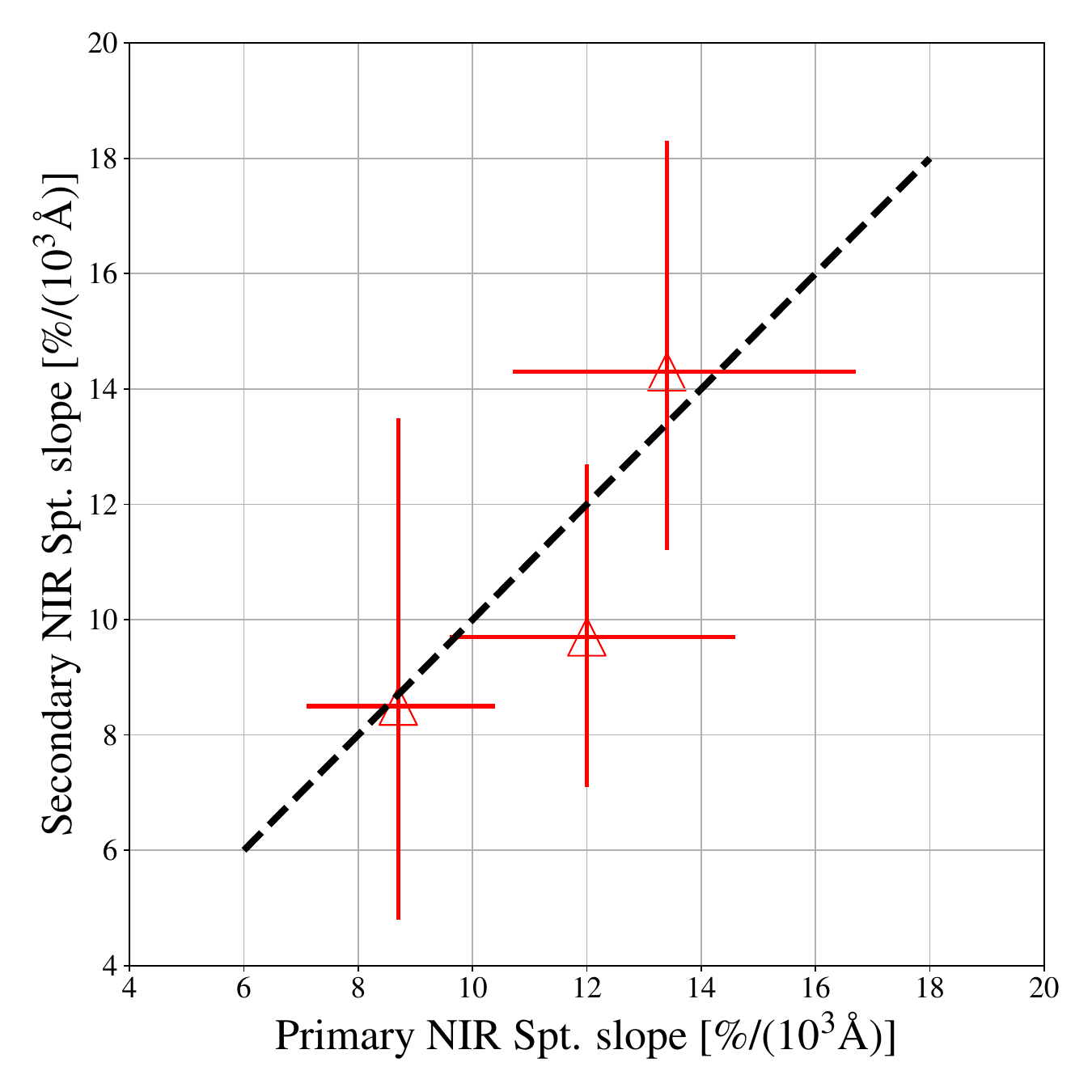}
\end{minipage}
\caption{ {\it Left:} Secondary vs. primary optical spectral slopes of TNBs. The blue circles correspond to previous $(V-I)$ color measurements from \citet{Benecchi:2009jo}, and the red triangles and magenta square correspond to new measurements derived from our $(g-r)$ and $(r-z)$ colors, respectively.
Spectral slope values for 505447~(2013~SQ99) derived from $(g-r)$ and $(r-z)$ colors are shown connected with a dotted line. 
{\it Right:} Secondary vs. primary near-infrared (622--1250\,nm) spectral slopes from this work. 
In both panels the dashed line indicates a 1:1 ratio, corresponding to perfect color equality of the binary components. 
None of the three binaries deviate from the color correlation in the optical and short near-infrared at the 2$\upsigma$ level, supporting similar spectra and surface composition of their components. } 
\label{fig:spt_slopes}
\end{figure}

In the left panel of Figure~\ref{fig:spt_slopes}, we plot the spectral slope of the primary component vs. that of the secondary in the optical wavelength range, derived from $(g-r)$ and $(r-z)$ colors. 
Spectral slope values computed from  $(V-I)$ color measurements in \citet{Benecchi:2009jo} are shown for comparison (here, the spectral slopes values were pulled directly from their paper).  
While these different colors do not probe the exact same wavelength range, the fact that the vast majority of TNOs exhibit linear spectral slopes in optical wavelengths makes it reasonable to compare these datasets.
The right panel of Figure~\ref{fig:spt_slopes} shows a similar plot in the short near-infrared, derived from $(r-J)$ color measurements. 
Because most TNOs show a spectral inflexion close to 1\,$\upmu$m, these measurements are shown in a different panel from the visible data.
In both wavelength ranges, none of the three binaries deviates at the $>$2$\upsigma$ level from the color correlation observed by \citet{Benecchi:2009jo} in the optical, implying that the binary pairs most likely share similar optical and short near-infrared colors, and therefore similar reflectance spectra and surface composition. 

The largest color difference between two binary components in our sample is observed for 505447~(2013~SQ99) and is $\Delta{(r-z)}=0.18$. 
505447 is a very red object that belongs to the dynamically quiescent population of cold classical objects \citep{Tegler:2000fx, Brown:2001fu}. 
Compared to the $g$ and $r$ photometric bands, the $z$ band was shown to probe a distinct spectral feature that is diagnostic of cold classical surfaces \citep{Pike:2017gf}. 
Testing the correlation of colors in the $(r-z)$ wavelength range would be another useful test of surface homogeneity. 
Here, the observed $(r-z)$ color difference between the components of 505447 is an inconclusive 2$\upsigma$ result. 
For comparison, the largest color difference in $(g-r)$ color is found for 506121~(2016~BP81): $\Delta{(g-r)}=0.11$, and is contained within the 1$\upsigma$ errorbars on the measurements. 

Compared to the other two binaries in our dataset, 511551~(2014~UD225) exhibits a rather large brightness ratio of its components. 
Specifically, assuming the two components have a similar albedo, the primary would be about 7 times more massive than the secondary. 
The similar colors of the components measured in this system suggests that the color homogeneity of TNO binaries remains valid for systems with such large mass ratios.


\begin{deluxetable}{c | c | c | ccc | ccc | ccc}[h!]
\tabletypesize{ \scriptsize}
\tablecaption{ \label{tab:binary_photometry}
Colors of the Binaries. }
\tablehead{\multicolumn{1}{c}{ Object } & \multicolumn{1}{c}{ Color } & \multicolumn{1}{c}{ $H_r^*$ } & \multicolumn{3}{c}{ $(g-r)$ } & \multicolumn{3}{c}{ $(r-z)$ } & \multicolumn{3}{c}{ $(r-J)$ } \\ 
\multicolumn{1}{c}{  } & \multicolumn{1}{c}{ class$^{*}$ } & \multicolumn{1}{l}{  }  & \multicolumn{1}{c}{ binary$^{*}$ } & \multicolumn{1}{c}{ primary } & \multicolumn{1}{c}{ secondary } & \multicolumn{1}{c}{ binary$^{*}$ } & \multicolumn{1}{c}{ primary } & \multicolumn{1}{c}{ secondary } & \multicolumn{1}{c}{ binary$^{*}$ } & \multicolumn{1}{c}{ primary } & \multicolumn{1}{c}{ secondary } } 
\startdata
505447             & red       & 6.45 & $0.98\pm0.03$  & $0.97\pm0.07$              & $0.99^{+0.09}_{-0.11}$ & $0.56\pm0.03$             & $0.63\pm0.07$            & $0.45\pm0.10$              &  $1.51\pm0.06$ & $1.54^{+0.08}_{-0.09}$ & $1.46^{+0.10}_{-0.11}$  \\
511551             & neutral & 6.55 & $0.74\pm0.02$  & $0.74\pm0.03$              & $0.74\pm0.13$              & --                                  & --                                  & --                                    &  $1.42\pm0.06$ & $1.42\pm0.06$              & $1.42^{+0.18}_{-0.16}$  \\
506121             & neutral & 6.58 & $0.59\pm0.03$  & $0.64^{+0.08}_{-0.10}$  & $0.53^{+0.09}_{-0.14}$ & --                                  & --                                  & --                                    &  $1.60\pm0.07$ & $1.59^{+0.10}_{-0.09}$ & $1.62^{+0.11}_{-0.10}$  \\
\enddata
\tablenotetext{}{For comparison, solar colors are $(g-r)$=0.45,  $(r-z)$=0.09 and $(r-J)$=0.98.}
\tablenotetext{}{${}^{*}$\citet{Schwamb:2019}.}
\end{deluxetable}

\begin{deluxetable}{ccccccccc}[h!]
\tabletypesize{ \scriptsize}
\tablecaption{ \label{tab:gradients}
Spectral Gradients between two wavelengths. }
\tablehead{
\multicolumn{1}{c}{ TNO } & \multicolumn{2}{c}{ 475--622\,nm  } & \multicolumn{2}{c}{ 622--905\,nm } & \multicolumn{2}{c}{ 622--1250\,nm } \\
\multicolumn{1}{c}{ Number } & \multicolumn{1}{c}{ Primary } & \multicolumn{1}{c}{ Secondary} & \multicolumn{1}{c}{ Primary } & \multicolumn{1}{c}{ Secondary} & \multicolumn{1}{c}{ Primary } & \multicolumn{1}{c}{ Secondary }}
\startdata
505447 & $32.5 ^{+ 4.1 }_{- 4.4 }$ & $34.1 ^{+ 5.2 }_{- 6.5 }$ & $28.1 ^{+ 5.4 }_{- 5.3 }$ & $15.5 ^{+ 6.2 }_{- 5.3 }$ & $12.0^{+2.6}_{-2.4}$ & $9.7 ^{+ 3.0 }_{- 2.6 }$ \\
511551 & $18.4 ^{+ 1.7 }_{- 1.7 }$ & $18.2 ^{+ 8.0 }_{- 8.2 }$ & -- & --& $8.7^{+1.7}_{-1.6}$ & $8.5 ^{+ 5.0 }_{- 3.7 }$ \\
506121 & $12.2 ^{+ 4.8 }_{- 6.7 }$ & $4.7 ^{+ 5.8 }_{- 9.0 }$ & -- & --& $13.4^{+3.3}_{-2.7}$ & $14.3 ^{+ 4.0 }_{- 3.1 }$ \\
\enddata 
\tablenotetext{}{ Spectral gradients are in units of $\%/(10^3 \AA)$ and relative to 550\,nm. 
A spectral slope of 0 corresponds to solar colors. }
\end{deluxetable}

\section{Discussion}
\label{sec:discussion}

The correlated colors of the TNB pairs across the full optical and near infrared spectral range advocates for similar reflectance spectra and hence surface compositions for these objects. 
While most TNO binaries studied so far are composed of equal-mass components, this feature appears to remain valid for systems with rather large mass ratios, such as 511551~(2014~UD225), where the primary is about 7 times more massive than the secondary (assuming they have a similar albedo).
Considering the hypothesis that TNO surfaces are primordial and indicative of their formation location \citep{Marsset:2019}, 
the surface equality of the binary pairs supports the two key assertions made by \citet{Benecchi:2009jo} about TNO formation and the early outer Solar System, namely: 
TNOs formed in a locally homogeneous and globally heterogeneous protoplanetary disk, 
and the formation of binaries must have happened early, before the violent dispersal of the disk that was driven by the migration of the gas giants, in order to explain the similar composition of their components. 
Taken together with other observed characteristics of the Kuiper belt binaries, namely: 
(1) their nearly equal mass components ($m_2/m_1>10\%$; \citealt{Noll:2008bj}), 
(2) their relatively wide mutual orbits \citep{Parker:2010}, 
(3) their almost unanimous prograde orbital distribution (with the exception of the widest systems; \citealt{Grundy2019}), and 
(4) their predominance in the TNO population \citep{Fraser:2017kh}, this finding provides further support to the hypothesis of binary formation through the early gravitational collapse of pebble clouds in the turbulent gas disk \citep{Nesvorny:2010da}. 
Only this scenario is currently able to account simultaneously for all of the abovementioned properties of the TNO binary population.
The predominance of binaries in the Kuiper belt was recently reinforced by the New Horizons flyby imagery of (486958)~Arrokoth (2014~MU69) \citep{Stern:2019}. 
This clearly showed Arrokoth to be a contact binary, implying a high binary fraction for systems below the current telescopic limits, at least in the cold classical population.


It should be noted, however, that several small TNOs are known to exhibit substantial color variability in the near-infrared that could relate to surface inhomogeneities induced by large impacts. 
This is the case, for instance, for the binary 26308 (1998~SM165) that exhibits 0.2 mag variability in the J--K spectral region \citep{Fraser:2015cx}. 
Two non-binary TNOs in the Col-OSSOS sample, 505448 (2013~SA100) and 2013~UN15, were also found to be spectrally variable in the optical and $J$-band \citep{Pike:2017gf, Schwamb:2019}.
Unless both components of such a system are rotationally locked and exhibit the same longitudinal color distributions, these highly spectrally variable systems must eventually break the color equality observed thus far.
The reason why some systems were not found to have differently colored components over a large range of sizes ($H_V$=4.0--8.1 in \citealt{Benecchi:2009jo}) is likely due to the low spectral variability of TNOs in the optical \citep{Fraser:2015cx} with respect to the precision of the observations. 
It appears more likely that such a signal has merely been hidden, though we note that a few systems presented by \citet{Benecchi:2009jo} exhibit 3$\upsigma$ deviations in the color difference of the primary and secondary. 

Like the sample in \citet{Benecchi:2009jo}, none of the three binaries analyzed in this work are known to be spectrally variable in the optical and near-infrared \citep{Schwamb:2019}. 
Although photometric sequences were acquired at a single epoch for each binary, their 1.7 to 3.0-hr duration allowed us to investigate but returned no hint of spectral variability. 
Moreover, the precision of measurements achieved in this work is insufficient to detect a 0.2 mag near-infrared color variation of the components. 
The required photometric precision needed to detect such variability will however certainly be achievable by future large ground-based facilities equipped with adaptive optics (e.g., the Extremely Large Telescope; \citealt{Gilmozzi:2007}, the Thirty Meter Telescope; \citealt{Sanders:2013}) and large space-based observatories (e.g., the James-Webb Space Telescope; \citealt{Gardner:2006}). 
Future surveys of the Kuiper belt will observe a small fraction of binary pairs with mismatched near infrared colors.

\section{Conclusion}

We report component-resolved photometry for three TNBs across the visible and near-infrared spectral range. 
These three objects present evidence that their components have similar colors in the visible and the near-infrared, in agreement with \citet{Benecchi:2009jo}'s finding that the components of a given TNB always share the same optical colors. 
This finding supports the inferences proposed by \citet{Benecchi:2009jo} that TNB pairs share the same surface composition, thereby advocating for early binary formation in a locally homogeneous, globally heterogeneous protoplanetary disk. 
However, considering that several small TNBs are known to exhibit substantial color variability in the near-infrared, 
we anticipate that the color equality of the binary will break down in some cases as future surveys increase the known binary sample. 

\section*{Acknowledgements}

The authors acknowledge the sacred nature of Maunakea, and appreciate the opportunity to observe from the mountain. 
This work is based on observations from the Large and Long Program GN-LP-1 (2014B through 2016B), obtained at the Gemini Observatory, which is operated by the Association of Universities for Research in Astronomy, Inc., under a cooperative agreement with the NSF on behalf of the Gemini partnership: the National Science Foundation (United States), the National Research Council (Canada), CONICYT (Chile), Ministerio de Ciencia, Tecnolog\'{i}a e Innovaci\'{o}n Productiva (Argentina), and Minist\'{e}rio da Ci\^{e}ncia, Tecnologia e Inova\c{c}\~{a}o (Brazil). 
We thank the staff at Gemini North for their dedicated support of the Col-OSSOS program. 
Data was processed using the Gemini IRAF package. 
STSDAS and PyRAF are products of the Space Telescope Science Institute, which is operated by AURA for NASA.
This research used the facilities of the Canadian Astronomy Data Centre operated by the National Research Council of Canada with the support of the Canadian Space Agency. 
M.M. was supported by the National Aeronautics and Space Administration under Grant No. 80NSSC18K0849 issued through the Planetary Astronomy Program.
B.J.G. and J.J.K. acknowledge funding from NSERC Canada. 
J.-M.P. acknowledges funding from PNP-INSU, France. 
K.V. acknowledges funding from NASA (grants NNX15AH59G and 80NSSC19K0785). 

\facilities{Gemini:Gillett (GMOSN, NIRI), CFHT (MegaCam)}
\software{Astropy \citep{Collaboration:2013cd}, Matplotlib \citep{Hunter2007}, NumPy \citep{vanderWalt:2011dp}, SciPy \citep{vanderWalt:2011dp}, Sextractor \citep{Bertin:1996hf}, pyraf, TRIPPy \citep{Fraser:2016bd}, emcee \citep{ForemanMackey:2013}, IRAF \citep{Tody:1986df}, Gemini IRAF package, STSDAS}

\bibliographystyle{apj}
\bibliography{references} 

\end{document}